# Intrusion Detection Using Cost-Sensitive Classification

Aikaterini Mitrokotsa[1*], Christos Dimitrakakis[2] and Christos Douligeris[3]

[1] Computer Systems Group,Vrije University Amsterdam,
De Boelelaan 1081a, 1081 HV Amsterdam, Netherlands
mitrokat@unipi.gr
[2] Chair of Information Technology, University of Leoben,
Erzherzog-Johann-Strasse 3, A-8700 Leoben, Austria,
christos.dimitrakakis@mu-leoben.at
[3] Department of Informatics, University of Piraeus,
Karaoli & Dimitriou Str. 80, 18534 Piraeus, Greece,
cdoulig@unipi.gr

**Abstract.** Intrusion Detection is an invaluable part of computer networks defense. An important consideration is the fact that raising false alarms carries a significantly lower cost than not detecting attacks. For this reason, we examine how cost-sensitive classification methods can be used in Intrusion Detection systems. The performance of the approach is evaluated under different experimental conditions, cost matrices and different classification models, in terms of expected cost, as well as detection and false alarm rates. We find that even under unfavourable conditions, cost-sensitive classification can improve performance significantly, if only slightly.

**Keywords:** Intrusion Detection, Cost-Sensitive Classification.

# 1 Introduction

Standard classification problems require making the classification decision that minimizes the probability of error. However, for many problem domains, the requirement is not merely to predict the most probable class label, since different types of errors carry different costs. Instances of such problems include authentication, where the cost of allowing unauthorized access can be much greater than that of wrongly denying access to authorized individuals, and intrusion detection, where raising false alarms has a substantially lower cost than allowing an undetected intrusion. In such cases, it is preferable to make the classification decision that has minimum expected cost, rather than that with the lowest error probability.

While there has been an extensive body of work in this field, particularly in the domain of optimal statistical decisions (see for example [1] for an overview), this has been largely ignored in the domain of intrusion detection. We are currently aware of two other papers ([2], [3]) dealing with cost-sensitive intrusion detection, both using a wrapper algorithm (MetaCost [4] and Weighted [5] respectively) together with RIPPER [6]. Although both papers report results on the KDD database, neither does so for the given cost matrix, so direct comparisons between the statistical models employed herein and the wrapper algorithms are not possible. This paper attempts to answer some very basic questions about cost-sensitive classification. Firstly, to what extent must the test distribution match the training data distribution. Secondly, for the dataset used, are some methods consistently better than others, or is there some variability and why. Finally, to what extent does the false alarm rate grow when the cost of missed attacks rises with respect to the cost of false alarms.

The next section discusses how to use classification methods that can be readily embedded in the formal optimal statistical decision framework in order to create intrusion detection systems that will be effective in minimizing the expected cost of their operation and analyses the relationship between cost matrices and the desired trade-off between detection and false alarm rates. Finally, it gives a brief introduction to the classification models used. Section 3 outlines the experiments performed and we conclude with a discussion on the significance of the results and on future research directions.

## 2 Cost sensitive classification

Given a specification of costs for correct and incorrect predictions, the class decision should be the one that leads to the lowest expected cost, where the expectation is computed using the conditional probability of each class given the example, according to our model[†]. More formally, for a set $\Omega$ of $k$ classes let a $k \times k$ matrix $C$ such that $C(i, j)$ is the expected cost of predicting class $i$ when the true class is $j$. If $i = j$ then the decision is correct, while if $i \neq j$ the decision is incorrect. Furthermore, let $Y$, $H$ be random variables denoting the actual and hypothesized class labels. For any observations $x \in X$ the optimal decision will be the class $i$ that minimizes a loss function equal to the expected cost

$$L(x,i) = E[C \mid X = x, H = i] \equiv \sum_{j \in \Omega} P(Y = j \mid X = x) \cdot C(i.j) \qquad (1)$$

where $P(Y \mid X)$ denotes the conditional distribution of class labels given an observation, according to our model. In this framework, all that is necessary is a model that can estimate this probability. The cost-sensitive decision-making function $f: S \to \Omega$ would then simply chose the decision $i$ that minimises the expected cost given the decision and the example[‡]. More formally,

$$f(x) = \arg\min_{i \in \Omega} L(x,i) \qquad (2)$$

The form of the cost matrix $C$ will depend on the actual application. In general, it is reasonable to choose the diagonal entries equal to zero, i.e. $C(i, j) = 0$ for $i = j$, since correct classification normally incurs no cost. The other entries specify the cost of incorrectly misclassifying an example of class $j$ as belonging to class $i$. They should be non-negative if the diagonal is zero, i.e. $C(i, j) \geq 0$ for $i \neq j$. Note that when this is equal to 1, the cost measure is the same as the classification error measure.

---

[†] The implicit dependency on some model m can be made explicit by conditioning everything on the model. Then the expectation would be written $E[C \mid x, f, m]$ and the conditional class probability $P(y \mid x, m)$.

[‡] Which of course is not necessarily identical to the decision with the minimum error probability. Furthermore, this framework is easily extensible to the case where the set of decisions differs from the set of class labels.

## 2.1 Choice of the cost matrix

As an example, consider a cost matrix *C* for two classes, positive and negative. The cost of a false positive is *C(2, 1)*, while that of a false negative is *C(1, 2)* and we can set *C(1, 1) = C(2, 2) = 0*, i.e. a correct classification will have no cost. For intrusion detection applications, it is common to refer to attacks as positive and normal instances as negative example. Furthermore, the occurrence of false negatives (FN) is usually considered a worse kind of error than that of a false positive (FP), thus the matrix C should reflect that, by having *C(1, 2) ≥ C(2, 1)*. In some cases, such as in some benchmark databases for intrusion detection, the cost matrix is given, while in others it must be chosen by the user.

## 2.2 Algorithmic comparisons and alternative quality metrics

When comparisons are made between algorithms, it is important to use the same measure of quality for all of them. A common measure of quality is the empirical value of the expected value of the cost C measured over an independent test set D,

$$\hat{E}(C \mid D) = \frac{1}{|D|} \sum_{d \in D} C(f(x_d), y_d), \qquad (3)$$

where $d \equiv (x_d, y_d)$. Whenever such a cost matrix is set as the evaluation metric in a benchmark database, then it is preferable to use it. However, it is important to note that in much of the literature, the following pair of measures is used instead. The *Detection Rate (DR)*, and the *False Alarm (FA)* rate

$$DR = \frac{TP}{TP + TN}, \quad FA = \frac{FP}{TN + FP} \qquad (4)$$

where *TP, TN, FP, FN,* denote the number of true *(T)* and false *(F)* positives and negatives respectively. The aim would be to reduce *FA* rate, while at the same time increasing *DR*. Since this is not usually possible, a trade-off between the two quantities is often sought instead. While such a trade-off may be automatically accomplished through the use of an appropriate cost matrix[§], in this paper we will only use these quantities as a secondary alternative comparison metric.

---

[§] Let the expected cost be E[C] = qP(H = 1|C = 2)P(C = 2) +r P(H = 2|C = 1)P(C =1) = q(1 - DR)P(C = 2) + rFA P(C = 1), where 2 denotes a positive example. Setting r = 1/P(C = 1) and q = k/P(C = 2) we obtain a cost function minimiz-

## 2.3 Models

As mentioned in the beginning of Section 2, the computation of class probabilities is model-dependent. Ideally one would assume a Bayesian viewpoint and consider a distribution over all possible models in a set of models, however in this case we will only consider point distributions in model space, i.e. a single parameter vector in parameter space. While this can cause problems with overfitting, we will use frequentist model selection methods to avoid this potential pitfall. These are described further in Section 3.2. The rest of this section gives a brief overview of the two models used in this work, the multilayer perceptron (MLP) and the Gaussian mixture model (GMM). A specific instance of an MLP can be viewed simply as a function $g: S \rightarrow \Omega$, where $g$ can be further defined as a composition of other functions $z_i: S \rightarrow Z$. In most cases of interest, this decomposition can be written as $g(x) = K(w'z(x))$, with $x \in S$, $w$ being a parameter vector, while $K$ is a particular kernel and the function $z(x) = [z_1(x), z_2(x), ...]$ is referred to as the *hidden layer*. For each of those, we have $z_i(x) = K_i(v_i'x)$, where each $v_i$ is a parameter vector, $V = [v_1, v_2, ...]$ is the parameter matrix of the hidden layer and finally $K_i$ is an arbitrary kernel. For this particular application wish to use an MLP $m$ as a model for the conditional class probability given the observations, i.e.

$$P(Y = y | X = x, M = m), \qquad y = g(x), \tag{5}$$

for which reason we are using a sigmoid kernel for $K$. In the experiments we shall be employing a hyperbolic tangent as the kernel for the hidden layer, when there is one. The case where there is no hidden layer is equivalent to $z_i = x_i$ and corresponds to the *linear* model. The GMM, the second model under consideration, will be used to model the conditional observation density for each class, i.e.

$$P(X = x | Y = y, M = m) \tag{6}$$

This can be achieved simply by using a separate set of mixtures $U_y$ for modeling the observation density of each class $y$. Then, for a given class y the density at each point $x$ is calculated by marginalizing over the mixture components $u \in U_y$, for the class, dropping the dependency on $m$ for simplicity.

---

ing FA - kDR, with *k* being a free parameter specifying the trade-off we are interested in.

$$P(X = x \mid Y = y) = \sum_u P(X = x \mid U = u)P(U = u \mid Y = y) \qquad (7)$$

Note that the likelihood function $p(X=x|U=u)$ will have Gaussian form, with parameters $\Sigma_u$, the covariance matrix and $\mu_u$ the mean vector, while $P(U=u|Y=y)$ will be another parameter, the component weight[**]. Finally, we must separately estimate $P(Y = y)$ from the data, thus obtaining the conditional class probability given the observations

$$P(Y = y \mid X = x) = \frac{1}{Z} P(X = x \mid Y = y) P(Y = y) \qquad (8)$$

where $Z = \sum_{j \in \Omega} P(X = x \mid Y = y) P(Y = j)$ does not depend on $y$ and where we have again dropped the dependency on $m$.
The conditional class probabilities from either (7) or (8), depending on the model, can then be plugged into (1), for calculating the decision function (2).

## 3 Experiments

In order to examine the effectiveness of the proposed approach, we conducted a series of experiments under varying conditions. In our experiments we performed comparisons in terms of the weighted cost defined in equation (3) using four different models: the MLP, Linear, GMM with diagonal covariance matrixes and Naïve Bayes models (GMM with a single Gaussian). It was expected that using the cost matrix to make decisions would result in a lower cost than when not doing so, even if the models' class probability estimates are not very accurate. A particularly interesting question was how the divergence between the training and testing data distributions affects the measured cost, for a fixed cost matrix. We furthermore investigated how the false alarm and detection rates change when we vary the relative cost of false alarms and false negatives.

### 3.1 Databases

We performed our experiments on the KDD database [7], using the 10% KDD dataset for training and cross-validation. The KDD dataset include

---
[**] Since we use separate mixture components for each class, $P(U=u|Y=y)=0$, when $u \notin U_y$, which also allows us to drop the dependency on $y$ in the likelihood function.

four types of attacks Denial of Service (DoS), Remote to Local (R2L), User to Root (U2R) and Probe.

**Denial of Service (DoS):** The main aim of a DoS attack is the disruption of services by attempting to limit access to a machine or service. Examples are back, land pod teardrop, smurf and neptnune.

**Remote to Local (R2L):** In a remote to local attack the attacker gains unauthorized local access from a remote machine and exploits this access in order to send packets over the network. Examples are Ftp_write, Guess passwd, Imap, warezclient, warezmaster, phf, spy and multihop.

**User to Root (U2R):** In U2R the attacker gains unauthorized access to local super user (root) privileges. Examples are Loadmodule, Perl, rookit and buffer overflow.

**Probe:** the attacker scans a network in order to find vulnerabilities requires little technical expertise. Examples are ipsweep, nmap, portsweep and satan.

Furthermore we performed two evaluations. For the first evaluation, we used the standard test KDD dataset, which includes 311,029 connections, including 17 types of attacks which are never observed in the training dataset. More specifically, in the full test data there are 4 new U2R attacks that correspond to the 92.90% (189/228) of the U2R class, 7 new R2L attacks that correspond to 63% (10196/16189) of R2L class in that dataset, 4 new DoS attacks that correspond to 2.85% (6555/229853) of the DoS class and 2 new types of Probe attacks that correspond to 42.94% (1789/4166) of all the Probe attacks. For this reason, for our second evaluation we used a version of this dataset which does not include these novel attacks. In both cases, the probability distribution of the test datasets is not the same as that of the training dataset.

The datasets are summarised in Table 1.

**Table 1.** Proportion of attack and normal connections for training and testing datasets.

| Datasets | Probe | DoS | R2L | U2R | Total Attacks | Total Normal |
|---|---|---|---|---|---|---|
| 10% KDD Dataset | 4107 | 391458 | 1126 | 52 | 396743 | 97278 |
| Test Dataset 1 | 4166 | 229853 | 16189 | 228 | 250436 | 60593 |
| Test Dataset 2 | 2377 | 223298 | 5993 | 39 | 231707 | 60593 |

### 3.2 Technical details

In order to select the best parameters for each model we performed 10-fold cross validation. For each MLP model we tuned three parameters, the *learning rate (η)*, the number of *iterations (T)* and the number of *hidden units (n_h)*. Keeping stable the *n_h* (equal to 0) we selected the appropriate *η* among values that range between 0.0001 and 0.1 with step 0.1 and the appropriate *T* selecting among 10, 100, 500 and 1000. For the selection of the appropriate *n_h*, having selected the appropriate *η* and the appropriate of *T*, we examined various values for *n_h* and selected the best among 10, 20, 40, 60, 80, 100, 120, 140, 160, 320. We additionally, used the MLP model with no *hidden units* as a *Linear* model.

For the GMM model we also tuned three parameters the *threshold (θ)*, the number of *iterations (T)* and the number of *Gaussian Mixtures (n_g)*. Keeping stable the *n_g* (equal to 20) we selected the appropriate *θ* among values that range between 0.0001 and 0.1 with step 0.1 and the appropriate *T* among 25, 100, 500 and 1000. For the selection of the appropriate *n_g*, after selecting the appropriate *θ* and the appropriate *T*, we examined various values for the *n_g* and the selected the best among 10, 20, 40, 60, 80, 100, 120, 140, 160, 320. We additionally, used the GMM model with one Mixture component as a Naïve Bayes model.

We have used the cost matrix defined for the KDD 1999 Dataset [8], which is shown in Table 2. We have also defined an arbitrary table in order to examine how the measured cost changes when the relative cost for the misclassification of attack versus normal connection increases. Thus, we define a 5x5 cost matrix $A$ where $A(j,1)=α$, and $A(1,i)=1$ for $j=2,3,4,5$ and $i=2,3,4,5$. Also $A(i,j)=0$, for $i=2,3,4,5$ and $j=2,3,4,5$, $A(1,1)=0$ and $a$ take values between 1 and 10 with step 1. Since the fields of the KDD dataset include discrete and continuous values, we represent the discrete values using one hot encoding. Furthermore, to ensure a good behaviour of all training algorithms we have normalized all the datasets to zero mean and unit variance.

**Table 2.** Cost matrix for the KDD 99 dataset.

| Predicted / Actual | Normal | Probe | DoS | U2R | R2L |
|---|---|---|---|---|---|
| Normal | 0 | 1 | 2 | 2 | 2 |
| Probe  | 1 | 0 | 2 | 2 | 2 |
| DoS    | 2 | 1 | 0 | 2 | 2 |
| U2R    | 3 | 2 | 2 | 0 | 2 |
| R2L    | 4 | 2 | 2 | 2 | 0 |

### 3.3 Results

We have evaluated each algorithm both with and without the use of a cost matrix for making decisions. Table 3 shows the results for Dataset 1, while Table 4 for Dataset 2. In both cases, μ is the expected cost, while low and high are the boundary values of the 99% confidence interval. The latter was estimated using 1000 bootstrap [9] samples of the test datasets.

It is clear that the empirical average cost for Dataset 1 is much higher than the corresponding cost for Dataset 2. This was expected, since Dataset 1 includes types of attacks that were not included in the training data. It is also evident that the Linear and GMM classifiers both achieve better results when we are using the cost matrix to make decisions. However, this was not the case for either the MLP or the Naive Bayes classifier. The latter's failure could be attributed to the fact that the Naive classifier assumes that all features of the training Dataset are independent. The reason for the behaviour of the MLP is not entirely clear. One possibility is that the probabilities that the model outputs do not accurately represent the uncertainty of classification, i.e. the classifier is 'too confident' due to the maximum likelihood training. However, we also note that nevertheless the MLP model performed as well as the weighted GMM model. This hypothesis is consistent with the fact that the MLP nevertheless performed as well as the weighted GMM model in Dataset 1, but significantly worse in Dataset 2.

We have performed a series of experiments for the MLP classifier, since this one presents the lower classification error, for the arbitrary cost matrix described in section 3.2. We examined how the performance of the MLP changes when we increase the cost (α) of false positives relative to that of false negatives. Again, we used the boostrap methodology to obtain confidence intervals for the results.

**Table 3.** Expected cost (μ) and boundary values (Low, High) with confidence 99% for Testing Dataset 1 with and without the use of a cost matrix.

|  | Without Cost | | | With Cost | | |
|---|---|---|---|---|---|---|
|  | Low | μ | High | Low | μ | High |
| MLP | 0.2384 | 0.2427 | 0.2472 | 0.2390 | 0.2431 | 0.2476 |
| Linear | 0.2425 | 0.2467 | 0.2511 | 0.2414 | 0.2452 | 0.2489 |
| GMM | 0.2497 | 0.2538 | 0.2578 | 0.2378 | 0.2420 | 0.2457 |
| Naïve Bayes | 0.3786 | 0.3829 | 0.3871 | 0.5304 | 0.5400 | 0.5353 |

**Table 4.** Expected cost (μ) and boundary values (Low, High) with confidence 99% for Testing Dataset 2 with and without the use of a cost matrix.

|  | Without Cost | With Cost |
|---|---|---|

|             | Low    | μ      | High   | Low    | μ      | High   |
|-------------|--------|--------|--------|--------|--------|--------|
| MLP         | 0.0711 | 0.0736 | 0.0759 | 0.0713 | 0.0739 | 0.0765 |
| Linear      | 0.0735 | 0.0761 | 0.0784 | 0.0716 | 0.074  | 0.0763 |
| GMM         | 0.0821 | 0.0845 | 0.0869 | 0.0686 | 0.071  | 0.0734 |
| Naïve Bayes | 0.2173 | 0.2204 | 0.2231 | 0.3733 | 0.3775 | 0.3817 |

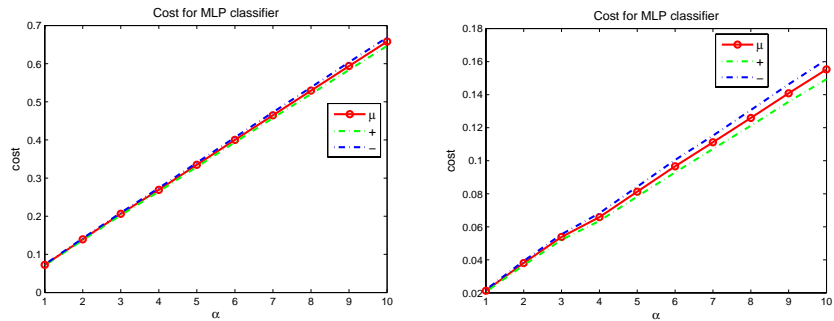

(a) Expected Cost for Testing Dataset 1    (b) Expected Cost for Testing Dataset 2

**Fig. 1.** Expected cost (μ) for Testing Datasets 1 (a) and 2 (b) respectively, with the cost of false negatives (attacks detected as normal) being equal to α, and the cost of false positives (false alarms) set equal to 1. The dashed lines (+,-) represent the bootstrap estimate of the 99% confidence intervals.

In Fig. 1 (a) and (b) the middle line (μ) represents the expected cost as it was estimated for the testing Datasets 1 and 2 respectively. The surrounding lines denote the 99% confidence interval of the expected cost as this was estimated from the bootstrap samples. From Fig. 1 (a) and (b), it is clear that the expected cost for Datasets 1 and 2 respectively, increases linearly with α, which indicates a good behavior of the classifier.

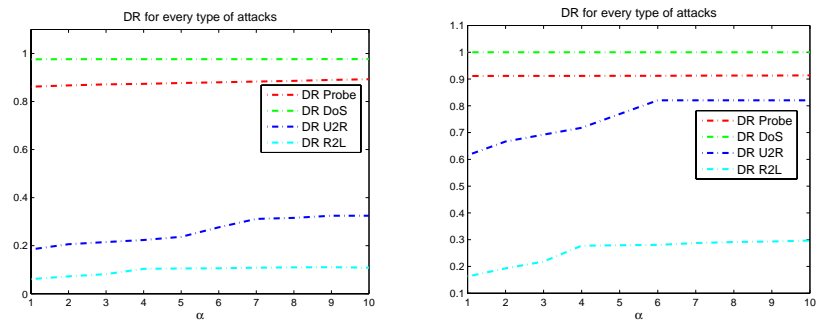

(a) Detection Rate for Testing Dataset 1    (b) Detection Rate for Testing Dataset 2

**Fig. 2.** The Detection Rate (DR) evaluated on Testing Datasets 1 (a) and 2 (b) respectively, with the cost of false negatives (attacks detected as normal) being equal to α, and the cost of false positives (false alarms) set equal to 1.

In Fig. 2 (a) and (b) we observe the detection rate for each type of attack for both testing datasets. The Detection Rate (DR) of all attacks is better for Dataset 2 while there is an increase of the Detection Rate for all type of attacks and both Datasets. While we observe a significant increase in the detection rate for the attacks in the training Dataset, this is not the case for the novel attacks, especially for the R2L attacks. The Detection Rate for DoS and Probe attacks presents only a slight increase for both cases. While for U2R attacks and Dataset 1 the Detection Rate ranges from 0.184 to 0.325 an increase of 14.1%. For U2R attacks and Dataset 2 we also observe a substantial increase of the Detection Rate from 0.615 to 0.82, an increase of 20%. For R2L attacks and Dataset 1 there is an increase of the Detection Rate from 0.06 to 0.108, an increase of 10.2%. For R2L attacks and Dataset 2 the Detection Rate ranges from 0.163 to 0.297, an increase of 13.4%.

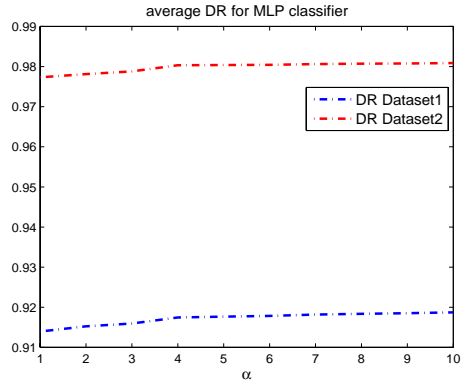

**Fig. 3.** The average DR for each Testing Dataset using the MLP classifier for various values of the cost of false positives relative to that of false negatives (α).

In Fig. 3 we observe that the average Detection Rate (DR) presents a slight increase from 0.913 to 0.919 for Dataset 1 and from 0.977 to 0.98 for Dataset 2. Fig. 4 depicts how the False Alarm (FA) rate is influenced by the increase of the cost of false positives relative to that of false negatives. We observe a slight increase from 0.016 to 0.018, thus the increase is of 0.2%. The False Alarm (FA) rate is the same for both Datasets since the number of normal connections is the same in both Datasets. Overall, the detection was increased significantly for the U2R (20%) and R2L (13.4%) attacks for Dataset 2, but *not* for novel attacks in these categories. In any case, the increase in false alarms was only 0.2%.

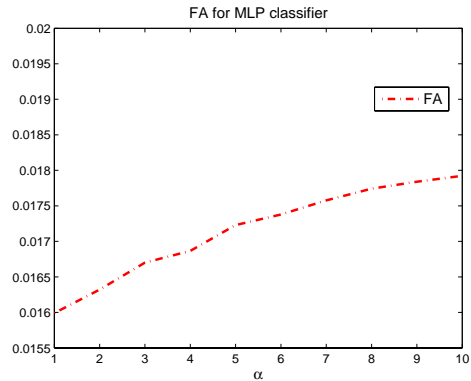

**Fig. 4.** The FA rate for each Testing Dataset using the MLP classifier for various values of the cost of false positives relative to that of false negatives ($\alpha$).

## 4 Discussion

The experimental results indicate that cost-sensitive classification methods using standard statistical classifiers to estimate class probabilities can behave quite well even in some cases where assumptions about the test data distribution are violated. This is not true for all methods tested. For the Naïve model, this can be explained by the fact that the assumption of feature independence cannot be maintained. However the behaviour of the MLP model is not as easy to explain. One possible explanation could have been that the predicted class probabilities by the MLP are more close to 1 than is warranted. However, Fig. 1-4 do not give support to this hypothesis. On the other hand, while the GMM's cost is significantly reduced when using weighted decisions, as can be seen by the lack of overlap between the confidence intervals, the MLP (whether weighted or not) is performing just as well as the weighted GMM in terms of cost. A possible explanation then is that the training and test data distributions are different enough for the generalization ability of a classifier to be more important than the use of the correct cost matrix.

Future work would have to further examine the relationship between distribution divergence and the use of cost matrices. An interesting ap-proach would be to use fully Bayesian methods, and also to perform more complete comparisons. A final point that we have only touched upon in the introduction is that the set of actions does not necessarily have to coincide with the set of classes. Then we would make decisions that minimise the expected cost of each decision. Examples of such decisions would be "Do nothing", "Call Administrator", "Block IP Address". Furthermore, we could even consider intrusion detection as a sequential decision making [1] problem, where each decision would not only depend upon the current observation, but on the history of observations and past decisions. This would not only make such systems more flexible, but could also reduce much of the future engineering performed in order to select what is the best time window in which to collect packet statistics.